\documentclass[fleqn,usenatbib]{mnras}

\usepackage{newtxtext,newtxmath}

\usepackage[T1]{fontenc}
\usepackage{ae,aecompl}

\usepackage{graphicx}	
\usepackage{amsmath}	
\usepackage{amssymb}	

\usepackage{color}

\title[Automated sky survey]{Automation of the AST3 optical sky survey from Dome~A, Antarctica}

\author[Ma, Hu, Shang et al.]{
\parbox{\textwidth}{
Bin Ma,$^{1}$
Yi Hu,$^{1}$
Zhaohui Shang,$^{1,2}$\thanks{E-mail: zshang@gmail.com}
Keliang Hu,$^{1}$
Yongjiang Wang,$^{1}$
Xu Yang,$^{1}$
Michael~C.~B.~Ashley,$^{3}$
Xiangyan Yuan,$^{4}$
and Lifan Wang$^{5}$
}
\\
$^{1}$National Astronomical Observatories, Chinese Academy 
of Sciences, Beijing 100101, China\\
$^{2}$Tianjin Astrophysics Center, 
Tianjin Normal University, Tianjin 300387, China\\
$^{3}${School of Physics, University of New South Wales, NSW 2052, Australia}\\
$^{4}${Nanjing Institute of Astronomical Optics and
 Technology, Nanjing 210042, China}\\
$^{5}${Purple Mountain Observatory, Nanjing 210008, China}\\
}

\date{Accepted XXX. Received YYY; in original form ZZZ}

\pubyear{2016}

\begin{document}
\label{firstpage}
\pagerange{\pageref{firstpage}--\pageref{lastpage}}
\maketitle

\begin{abstract}
The 0.5\,m Antarctic Survey Telescopes (AST3) were designed for time-domain
optical/infrared astronomy. They are located in Dome~A, Antarctica, where they can take advantage of the continuous dark time during winter.
Since the site is unattended in winter, everything for the
operation, from observing to data reduction, had to be fully
automated.  Here, we present a brief overview of the AST3 project and some of its unique characteristics due to its location in Antarctica. We summarise the 
various components of the survey, including the customized hardware and software, that make
complete automation possible.
\end{abstract}

\begin{keywords}
methods: observational -- 
methods: data analysis –-
methods: miscellaneous --
instrumentation: miscellaneous --
techniques: image processing --
software: development.
\end{keywords}


\section{Introduction}
\label{sec:intro}

Dome~A is the highest place on the ice cap of Antarctic plateau with an
elevation of about 4100\,m.  It is located at 77.56\degr\,E, 80.367\degr\,S, 
and polar nights can be enjoyed for months for astronomical observations. The uninterrupted darkness during winter is 
especially valuable for time-domain astronomy.  In addition, the atmospheric and geographic 
conditions at Dome~A have many benefits for astronomical
observations.

Dome~A was reached for the first time by the Chinese National Antarctic Research Expedition (CHINARE)
in 2005 and Kunlun
Station was later established at Dome~A in 2009.  Much work has been done on astronomical site-testing since 2007, confirming the dark sky background \citep{Zou10,Yang17},
favorable weather conditions \citep{Hu14,Hu19}, and the very high
atmospheric transmission in the terahertz bands \citep{Shi16}.

Moreover, a new study shows that the optical free-atmosphere seeing of
0\farcs31 (median), with periods as good as 0\farcs13, can be more frequently and easily obtained at Dome~A
\citep{Ma20} than at Dome~C \citep{Lawrence04,Agabi06,Aristidi09}.


In parallel with the site testing, astronomical observations have also
occurred.  The first generation optical telescope -- the
Chinese Small Telescope ARray \citep[CSTAR;][]{Yuan08,Zhou10} -- was
installed in 2008 and continuously monitored an area of about
20~deg$^2$ around the South Celestial Pole for three winters.  CSTAR
has four 14.5\,cm telescopes, but no moving parts, in order to minimize risks for an initial
experiment in the harsh environment.  CSTAR served as a pathfinder to test the practicality
of operating a telescope at Kunlun Station automatically
\citep{Zhou10}. 

Following the experience with CSTAR, a 2nd generation optical facility was developed: AST3.
The three Antarctic Survey Telescopes were designed for
time-domain astronomy as well as site testing.  Their two major
scientific drivers were a survey to discover supernova (SN), and an exoplanet search.
However, since AST3 are fully functional telescopes with pointing and
tracking, they can also serve as general-purpose optical/infrared telescopes.

Since Kunlun Station is not yet a winter-over station, all the 
instruments must be automated to work throughout the winter with no possibility of on-site human intervention. The CHINARE traverse team is only present at the station for a period of about 3 weeks each summer. During this time, the team must install and commission new telescopes, computers, and power systems, as well as retrieve data obtained during the previous year. Needless to say, commissioning a new telescope in bright sunshine at an altitude of 4100\,m and ambient temperatures of $-30^\circ$C to $-40^\circ$C is a challenging task.
Power and Iridium satellite communication are
provided by PLATO-A, an automated supporting facility
\citep{Ashley10a,Ashley10b}.

In this paper, we present details of the automation of the AST3 sky survey.  We
describe the AST3 telescope, the CCD camera, and key
hardware and software in Section~\ref{sec:instr}.  The observing operation and
data pipeline are described in Section~\ref{sec:operation}. 
A summary of the operation of the automated survey is presented in
Section~\ref{sec:summary}.

\section{Instruments and Control Software}
\label{sec:instr}

\subsection{The Telescopes}

Here, we provide background information on the key characteristics of the AST3 telescopes.
More details have been
presented in previous works \citep{Yuan10,Yuan12,Yuan14}.  

The AST3 telescope is a modified Schmidt design with a
transparent aspheric plate, an oblate primary mirror, and spherical
lens correctors in front of the focal plane \citep{Yuan12}.  This design achieves a
wide field of view (FOV) and a good image quality with 80 per cent of a point
source's energy enclosed within 1\arcsec.  Each telescope has an
entrance pupil diameter of 50\,cm and an f-ratio of 3.73.  The compact
design assists with easy transport and installation at Kunlun Station,
Dome~A. 

Each telescope was originally designed to have only one filter, however the
second telescope, AST3-2, was made with two-filters ($i$ and $R$-band), while the first telescope, AST3-1, was equipped with a single $i$-band
filter. The third telescope, AST3-3, is being fitted with a $K$-dark infrared filter and camera, and
has yet to be sent to Dome~A \citep{Li16,Burton16}.

The unique conditions on the Antarctic plateau mean that there is no dome provided or needed for the AST3 telescopes. The ambient temperature
varies from $-50\degr$C to $-80\degr$C during the observing season
\citep{Hu14,Hu19}, so low thermal expansion materials were chosen to
minimize the thermal effects on the optics and avoid frequent
focusing.  A dedicated telescope control computer ({\it TCC\/}) is
responsible for driving each telescope.  

\begin{figure}
	\includegraphics[width=\columnwidth]{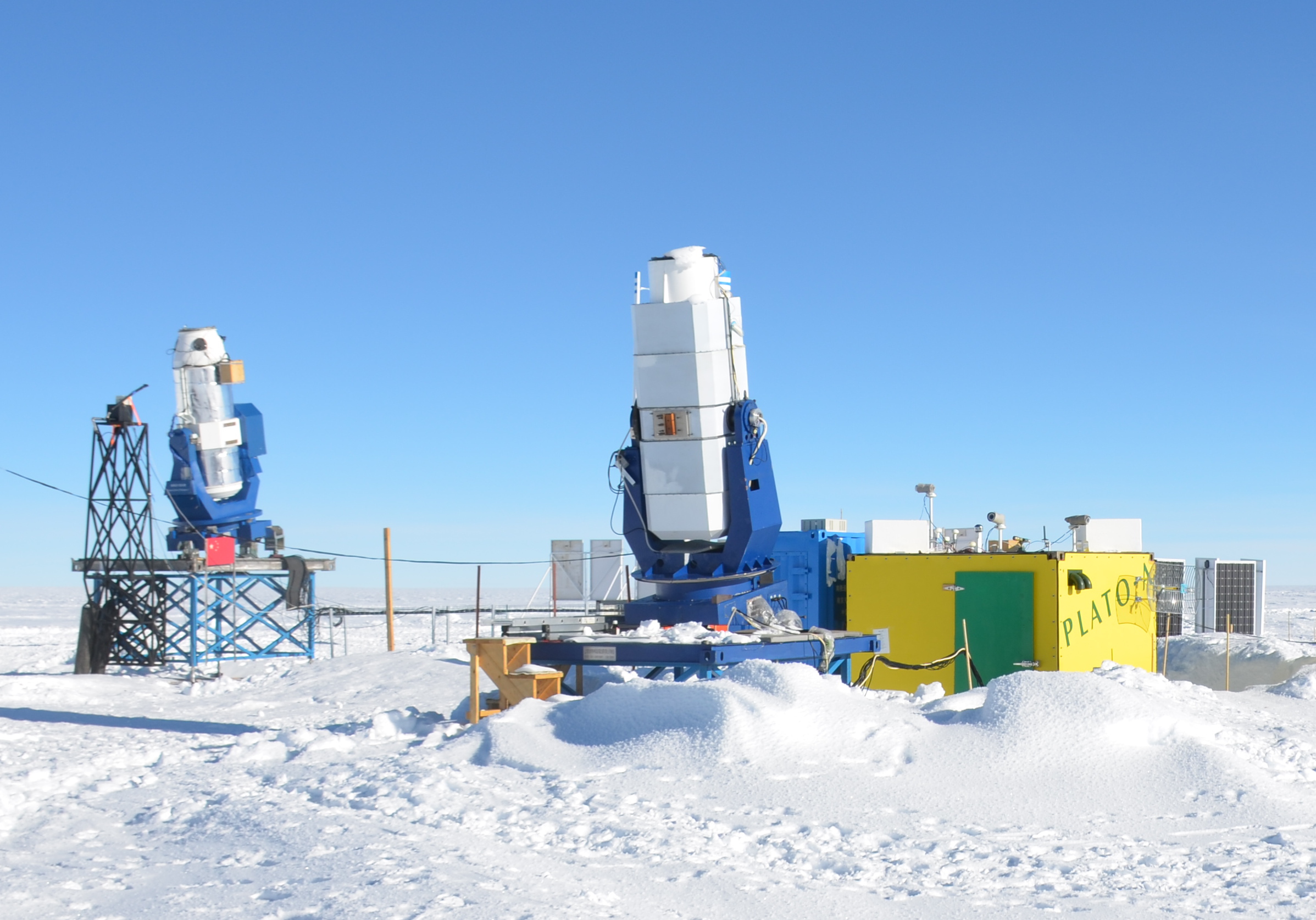}
    \caption{AST3-1 (right) and AST3-2 (left) at Dome~A.  Also shown is PLATO-A's
yellow instrument module, which houses the control computers, batteries, and Iridium communications.}
    \label{fig:telescope}
\end{figure}

AST3-1 and AST3-2 were installed at Kunlun Station in 2012 and 2015,
respectively (Figure~\ref{fig:telescope}).  
%
The telescopes are equatorially mounted, and a new method was
developed to enable the alignment of the optical system and the telescope 
without using stars during daytime at Dome~A \citep{Li14,Li15}.  As
a result, the misalignment of the polar axis is relatively small,
e.g., only 0.7\degr\
for AST3-1, compared to its large FOV of 4.26\,deg$^2$
(Sec.~\ref{sec:ccd}).  Moreover, with the help of a telescope pointing
model built with {\sc tpoint}
software\footnote{\url{http://www.tpointsw.uk/}} and data obtained during twilight -- some months after the traverse team had left Dome~A -- the pointing accuracy was improved to better than tens of
arcseconds.

\subsection{The CCD Camera}
\label{sec:ccd}

Each telescope is equipped with an STA1600FT CCD camera and Reflex
controller built by Semiconductor Technology Associates, Inc (STA)
\citep{Bred12}.  The Reflex controller is connected via a fiber optic cable to an
acquisition card in a computer (Sec.~\ref{sec:main}) within the PLATO-A instrument module.

STA provided with the CCD cameras control software called
{\it ast3\/}.  To conduct the sky survey
non-interactively, we developed our own software
(Sec.~\ref{sec:ast3suite}) to communicate with the {\it ast3} software through a
TCP network port.
STA also provided image acquisition software called {\it
reflexcapture\/} for recording images from the CCD chip. However,
this is an interactive program with a Graphic User Interface that is not
suitable for an automated survey.
We therefore also
developed our own image acquisition software
(Sec.~\ref{sec:ast3suite}). 

\begin{figure}
\begin{center}
	\includegraphics[width=7cm]{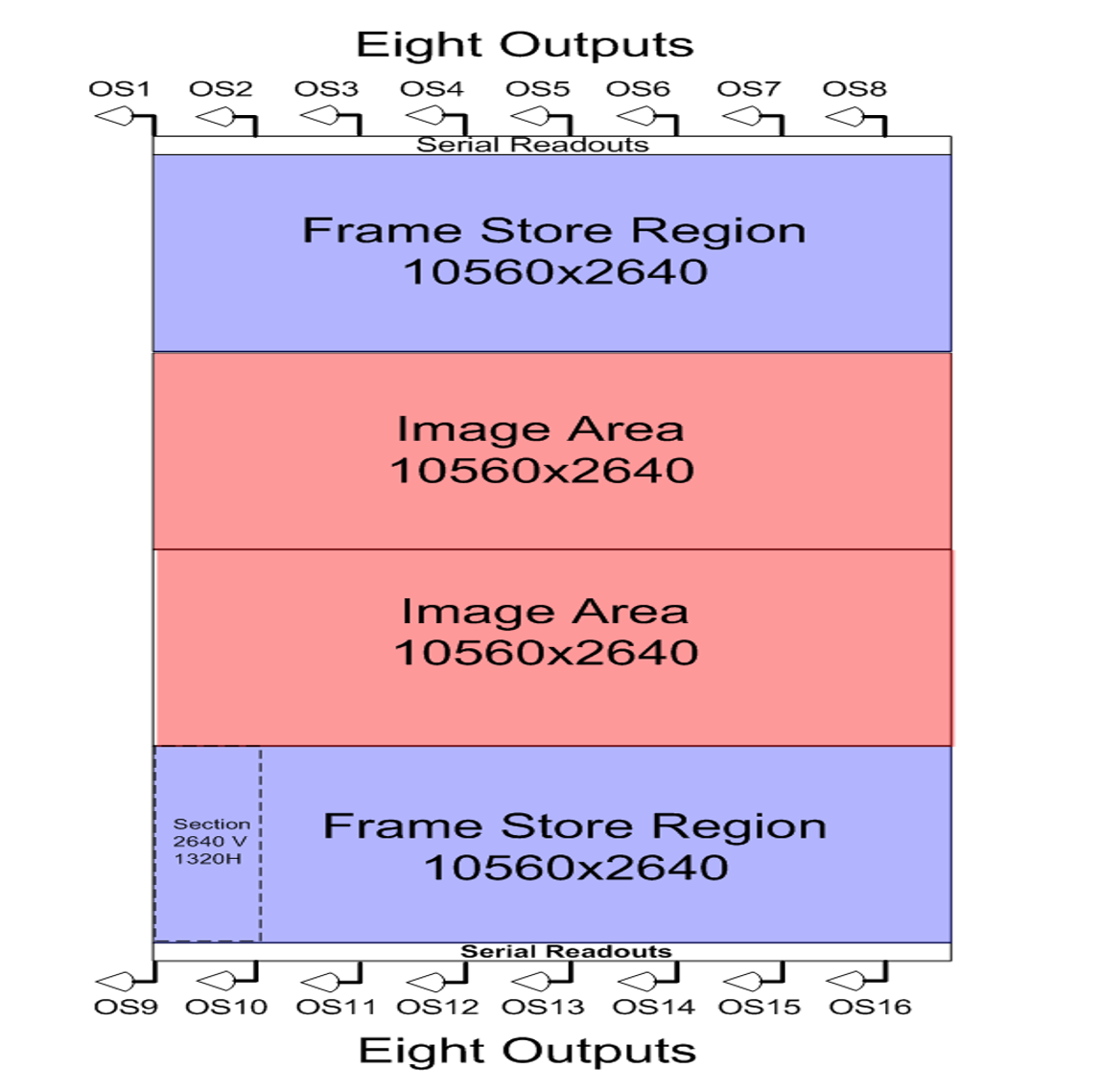}
    \caption{Illustration of the exposure and frame-transfer buffer areas
of the AST3 CCD (Adapted from a report from STA). }
    \label{fig:ccdframe}
\end{center}
\end{figure}

The large format 10k$\times$10k single-chip CCD has $10560\times10560$
9\micron\ pixels, resulting in a plate scale of 1\arcsec/pixel and
a FOV of 2.92$\degr \times 2.92\degr$.  
In order to avoid mechanical failure in cold weather, we
eliminated a shutter for the camera and instead operated the camera
in frame-transfer mode using only the central $10560\times5280$ pixels
as the exposing area -- the top and bottom $10560\times2640$ areas were shielded from the light, and used as
frame-transfer buffers (Fig.~\ref{fig:ccdframe}).  This gives a
final effective FOV of 2.92$\degr \times 1.46\degr$, i.e.
4.26\,deg$^2$.

To reduce the CCD readout time, there are 16 parallel readout channels with two readout
modes.  It takes 158.6\,ms (in fast mode) and 431.2\,ms (in slow mode) to
transfer the 10k$\times$5k image to the two buffers for readout.  This is
a small fraction of the readout time of 2.55 seconds
and 40.2 seconds for the fast (100\,kHz) and slow modes (1.6\,MHz),
respectively. The measured readout noise is about 4 e$^-$ for the slow
mode and 9--12~e$^-$ for the fast mode \citep{Ma12}.

\begin{figure} 
\includegraphics[width=\columnwidth]{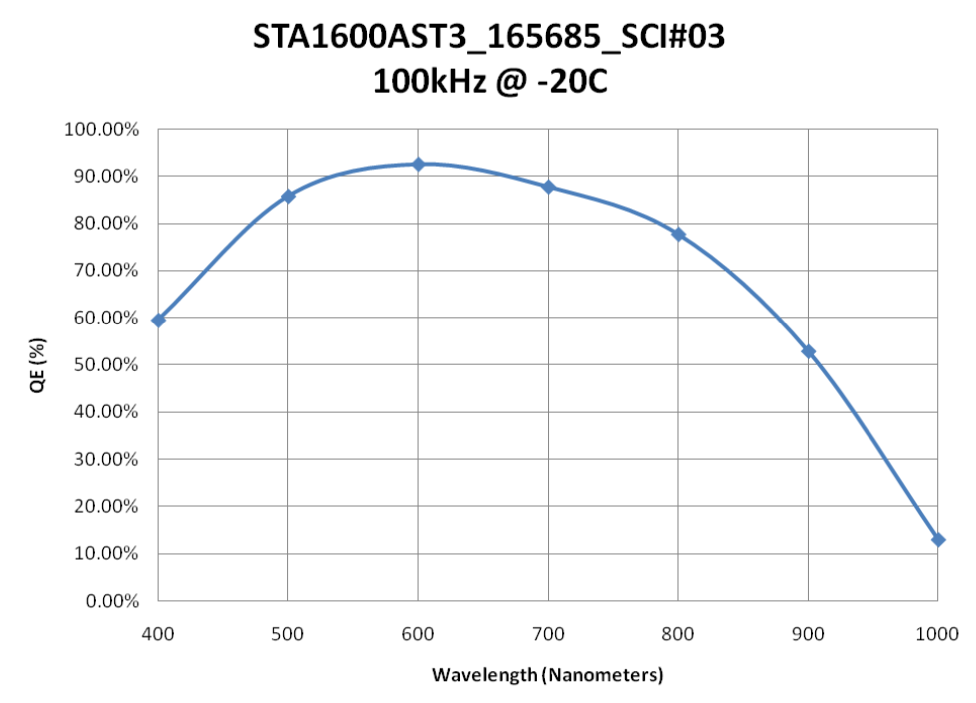}
\caption{Typical quantum efficiency of the AST3 CCD detector 
(Adapted from a report from STA).} 
\label{fig:qe}
\end{figure}

Figure~\ref{fig:qe}, provided by STA, shows the typical quantum
efficiency of the AST3 CCD camera.
The full well capacity for the camera is about 100,000\,e$^-$, and the
gain set for each channel is around 1.6--1.7\,e$^-$/ADU to match the
digital saturation level of 65,536 ADU for the 16-bit digitization.

\begin{figure}
	\includegraphics[width=\columnwidth]{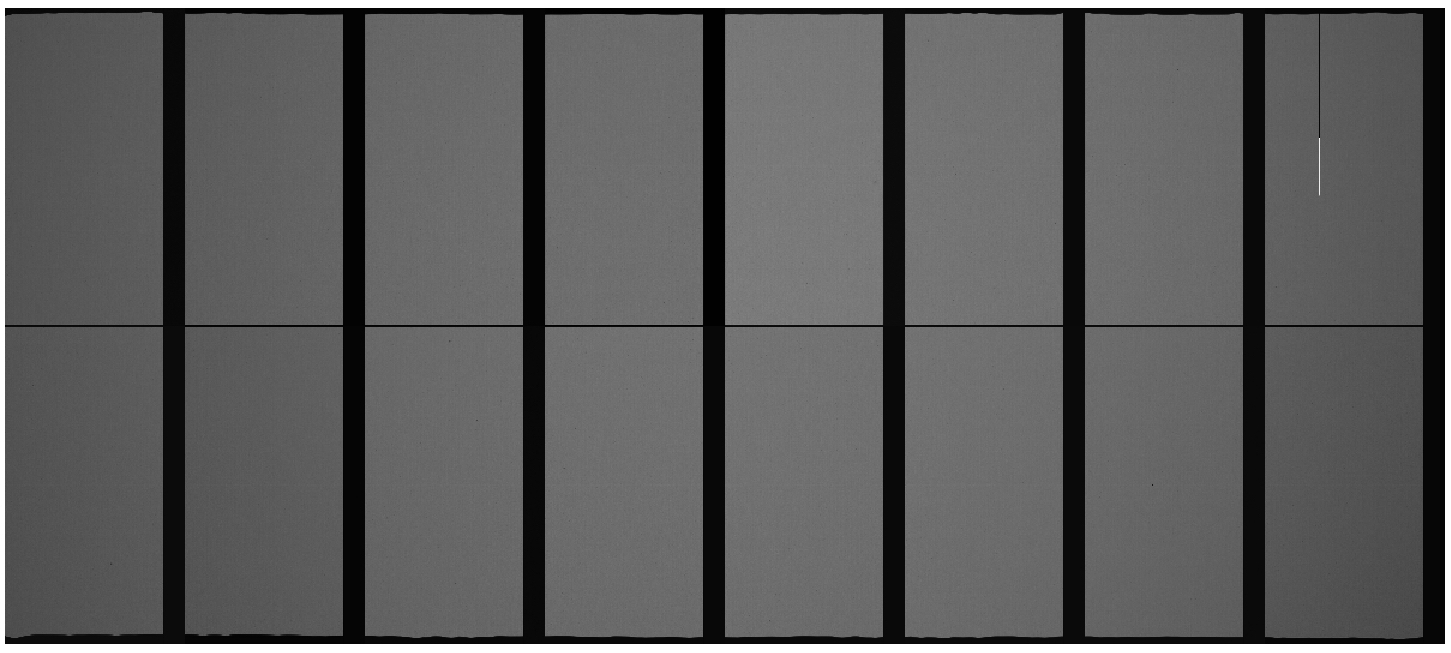}
    \caption{A 10k$\times$5k flat-field image for one CCD camera,
showing the 16 readout channels. The black gaps are 180-column horizontal and 10-line vertical
overscan regions for each channel -- there are no gaps on the sky.}
    \label{fig:flatframe} \end{figure}

Figure~\ref{fig:flatframe} shows a flat-field image taken with a
camera.  

By taking advantage of the low
ambient temperature at Dome~A, the CCD camera was cooled with a
single-stage thermoelectric cooler (TEC) that can generate a
temperature difference of $30\degr$C between the detector and the cryostat backplate.
%
The dark current at $-80\degr$C was estimated by STA to be about 0.0059  e$^-$/s,
based on a measurement at a higher temperature and extrapolating using a halving with every decrease of
$7\degr$C.  Our measurement
of the dark current in a cold chamber at $-80\degr$C was in the range 0.01--0.04 e$^-$/s, depending on channel number.

\begin{figure}
	\includegraphics[width=\columnwidth]{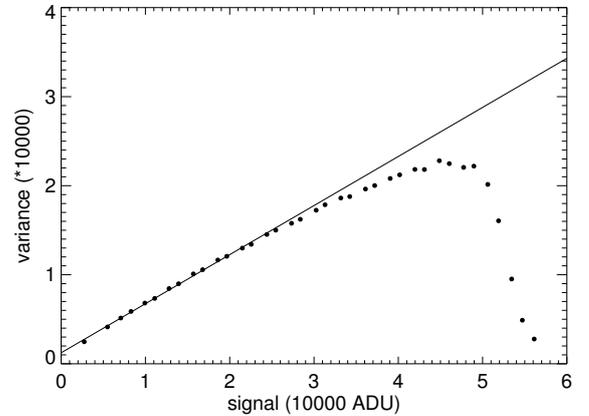}
    \caption{A typical photon transfer curve (one channel) of one AST3 CCD camera.} 
    \label{fig:ptc}
\end{figure}

Figure~\ref{fig:ptc} shows a typical photon transfer curve (PTC)
measured for one channel of one of our CCDs.  The PTC begins to become nonlinear at a
level at around 25000 ADU.  This has the effect of
redistributing electrons between pixels and results in a
changing PSF \citep{Ma14a} for bright stars.
However, this effect is negligible for sources in the linear regime.

\subsection{Control, Operation, and Data System}

We have built for each telescope a Control, Operation, and Data System (CODS) which is responsible for all the
functions of our automated sky survey.  Although we are able to
interact with CODS through the Iridium satellite network, our system was designed to
work fully automatically at Dome~A.

CODS consists of three sub-systems: the main control system, the data
storage array, and the pipeline system \citep{Shang12}. 
A customized software suite supports the operation of the entire system.
We describe each component in the following sections.

PLATO-A is able to provide an average of 1~kW
of power continuously for a year for all the instruments and computers.  To save power, CODS was
designed to minimize the power consumption of our computers and other components, while maintaining sufficient
computational power for the real-time pipeline.
Following testing we chose for the control computer a laptop configuration with an Intel i7-620M
CPU and a low temperature-rated ($-20\degr$C) motherboard with an Intel
HM55 chipset. The power needed per computer is less than 40\,W even when fully
loaded.

CODS has 6 identical computers, with two for each
sub-system for redundancy.  All internal communication is based on
TCP/IP on a local network except for the communication with the CCD
camera which is connected to CODS through a fiber optic cable.

\subsubsection{Main Control System}
\label{sec:main}

The main control system ({\it MAIN\/}) is the core of the
survey and consists of one computer and associated hard disks, and the
CCD acquisition card. {\it MAIN\/} communicates with the telescope and CCD
camera, acquires images, and sends them to the storage system and real-time
pipeline.  {\it MAIN\/} also monitors the status of all the instruments
and devices, obtains results from the pipeline, and sends alerts and
alarms through the Iridium network back to the control center at the
National Astronomical Observatories, Chinese Academy of Sciences
(NAOC) in Beijing (Sec.~\ref{sec:alarm}).

There is another identical {\it MAIN\/} computer for redundancy.
However, the CCD camera was designed to
communicate with one acquisition card in one computer via a fiber
optic cable.  To allow either of the {\it MAIN\/} computers to be able to control the camera, we developed a technique to split the
signal from the CCD controller into two fiber optic cables, one going to each acquisition card in
the two {\it MAIN\/}s.  This technique has proved to
work successfully, and was essential at one point following an acquisition card failure.  One consequence of this design is that the two identical {\it MAIN\/} systems cannot be
operated at the same time, but this is never desired or needed.

\subsubsection{Data Storage Array}
\label{sec:array}

\begin{figure}
	\includegraphics[width=\columnwidth]{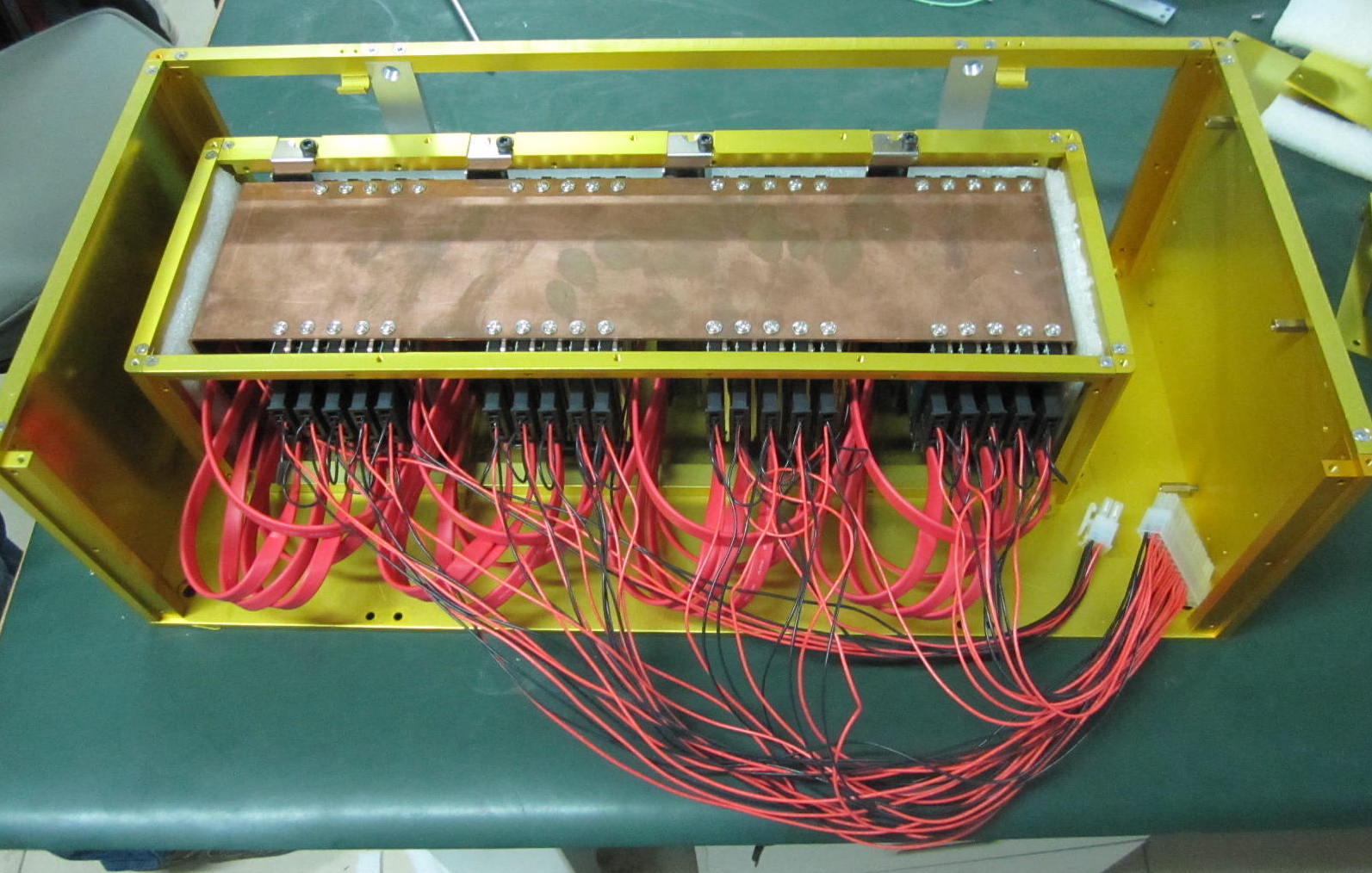}
    \caption{Our customized disk array built from 2.5-inch disks 
organized into four groups of five disks each.}
    \label{fig:array}
\end{figure}

Our data storage array ({\it ARRAY\/}) was fully customized as
there was no available product that met our requirements.  Each of
the two identical {\it ARRAY\/}s uses one dedicated computer to manage the
disks and image data. For reasons of cost, we used twenty
2.5-inch 500\,GB disks (Fig.~\ref{fig:array} ) to form an array with 10\,TB of capacity,
enough for the first year of operation. We added a PCIe-to-SATA
card which has four SATA ports, with each SATA port connected to
a SATA multiplier that can support five disks. 
By increasing the capacity of each single disk, we can easily increase
the storage without increasing the physical volume of {\it ARRAY\/}.  As new
technology developed, our second-generation {\it ARRAY\/} dropped the SATA
multiplier and used higher capacity (8--10\,TB) helium-filled 3.5-inch hard  disks. These disks can
better cope with the low air pressure at Dome~A, which is equivalent to an altitude of about 4500\,m, and is below the minimum operating pressure of most hard disks. Hard disks were preferred over solid-state disks (SSDs), since SSDs have worse temperature performance and in our experience are more likely to suffer complete data loss.

{\it ARRAY\/} was designed to work primarily in write mode, in the sense that the software normally only provides an interface for writing images.  {\it ARRAY\/} receives each
raw image from {\it MAIN\/} immediately after an exposure, and simply stores the image on the appropriate hard
disk.  
For reliability, disks are powered in pairs through two parallel
switches controlled by the computer, and usually only two working disks
are powered on at a time.  The disks are powered off once they are full, to
save energy and also to safeguard the data.  However, if needed, we can
remotely turn on any disk in {\it ARRAY\/} and read the raw data.

\subsubsection{Pipeline Computers and Supporting Software}
\label{sec:pipe}

The pipeline system ({\it PIPE\/}) is relatively simple and
is mainly responsible for real-time data processing.  Each of the two
identical {\it PIPE\/}s consists of a computer and its associated large-capacity hard disks used for
calibration images, templates, and temporary files as well as
photometry catalogs. The catalogs are also stored on {\it MAIN\/} as a  backup
(Sec.~\ref{sec:pipeline}).  Only one {\it PIPE\/} system is powered on at any time during the
observing season.

The data reduction is designed to run automatically with the
support of a {\it Pipeline Daemon\/} running on {\it PIPE\/} \citep{Yu14}.
This daemon triggers the AST3 pipeline (Sec.~\ref{sec:pipeline}) whenever
a new scientific image is received from {\it MAIN\/} or if the image queue
is not empty.  There is another daemon process running in parallel with
{\it Pipeline Daemon\/} for cross-protection and crash recovery -- if
either process crashes or times-out, the other will restart it.
In addition, Linux {\it crontab} entries provide
another layer of protection for the two parallel processes.
This design produced a highly reliable
real-time pipeline process.

To take advantage of multiple cores and hyper-threading,
each image was processed using software that was parallelized on a channel by channel basis.
{\it Pipeline Daemon\/} is able to take care
of multiple processes at the same time.

\subsubsection{AST3 Software Suite}
\label{sec:ast3suite}

The first version of our control software was based on a single daemon program, and had many limitations in practice. Our second version was called
{\it ast3suite} \citep{Hu16} and was sufficiently flexible to use for any robotic
observatory \citep{Hu18}.

\begin{figure*}
	\includegraphics[width=14cm]{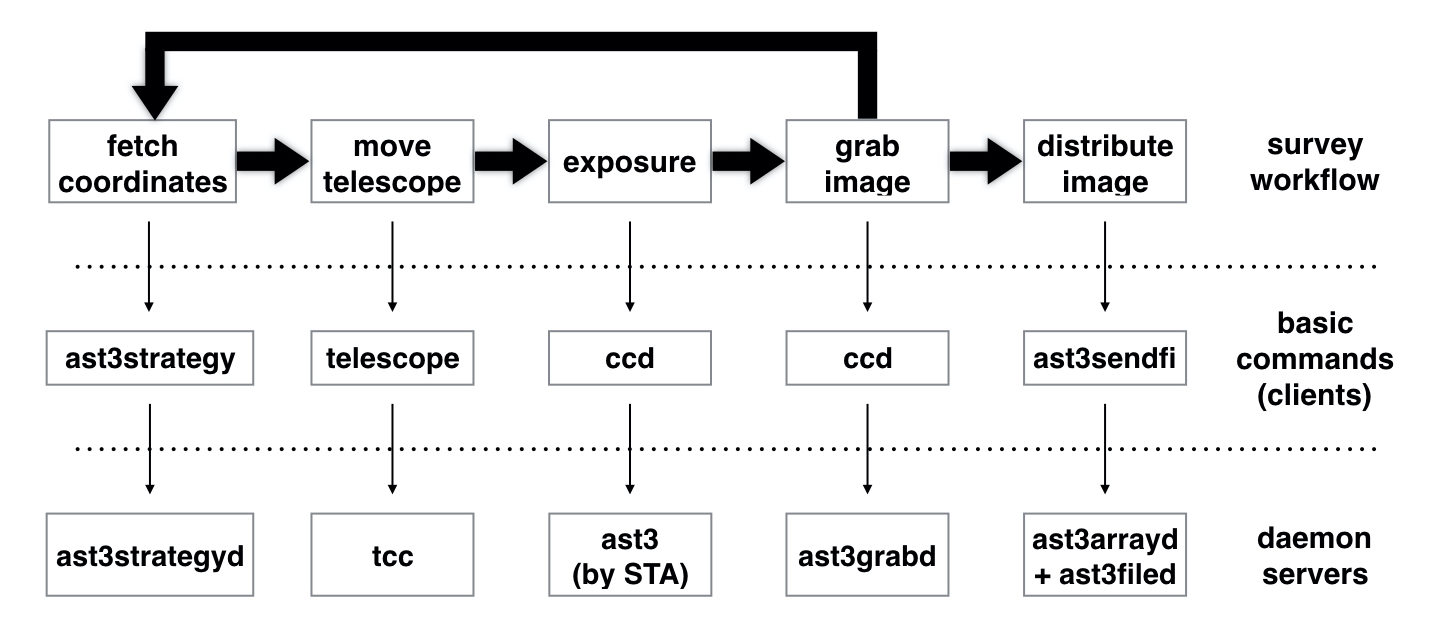}
    \caption{An example of the observation loop for our sky survey, based on {\it
ast3suite}.  The three levels of software comprising {\it ast3suite} are shown
(Sec.~\ref{sec:ast3suite}).}
    \label{fig:ast3suite}
\end{figure*}

Based on a client-server architecture, {\it ast3suite} is designed to
have three levels of software as shown in Fig.~\ref{fig:ast3suite}.  The bottom level has a number of
daemon servers, each of which directly operates on a single type of
device, such as a computer or a serial-device.
The middle level consists of clients that issue basic commands 
to the daemons.  The top level uses higher-level
languages and scripts to flexibly integrate the basic commands
in order to carry out more complicated tasks.

We are able to connect to {\it MAIN\/} over the Iridium satellite network from NAOC and control
all the instruments using the basic commands, giving us a flexible and
efficient system for engineering work and manual observations. However, for the sky survey it
was preferable to automate the operation for control on-site, since the Iridium connection was relatively slow and expensive and had a long latency.

\subsubsection{Cold Test and Field Test}
\label{sec:test}

All the hardware of CODS was extensively cold-tested down to
$-20\degr$C, the lowest temperature expected in PLATO-A's instrument module.
This gave us confidence in its ability to operate successfully at
Dome~A.  The CCD camera and controller have been tested down to
$-80\degr$C and $-55\degr$C, respectively. If the controllers are initially off, they must be warmed to at least $-55\degr$C before being turned on. Their minimum storage temperature is lower,
and they have been tested down to at least $-70\degr$C.

In order to better understand and optimize their overall performance,
AST3-2 and CODS were tested for more than four months over winter in
Mohe, China, where the temperature can drop to below $-40\degr$C.
Both the hardware and software of CODS worked as expected and
we conducted a real sky survey simulating the operation at Dome~A
without any human intervention.  The test demonstrated the reliability
of the system.

\section{Operation and Pipeline}
\label{sec:operation}

The AST3 sky survey was designed to be fully automated including the
observation and data reduction.  Due to the limited bandwidth to Dome~A all the software 
and calibration catalogs had to be prepared before
the actual survey began.  

\subsection{Observing Strategy}

In order to achieve an unattended, automated sky survey, and to utilize the
available observing time most efficiently, 
we carefully planned an observing strategy and implemented a
{\it survey scheduler}.  The scheduler software determines the best
sky field for the next observation by taking into account the survey areas, 
the fields that have been observed, the required
observing cadence, the telescope slew time, the airmass, and the sky background
based on the locations of the Sun and the Moon and the phase of the
Moon \citep{Liu18}.

The {\it scheduler} works with a target-list configuration file that 
is highly configurable for parameters such as observing fields,
exposure times and cadences.
Since the AST3 sky survey mostly focused on a SN survey and
exoplanet search, we pre-defined survey modes for these two cases based on their scientific requirements. We provided an additional
special mode for rapid follow-up of interesting transients
such as GRBs or gravitational wave events etc.

The SN survey mode was optimized for a large sky area of 2000\,deg$^2$
with a cadence of a half to a few days depending on the fraction of dark time during a 24 hour period.
The exoplanet search mode covered
relatively small areas, with 10--20 fields per target region, and a
cadence of tens of minutes with short exposure times so as not to saturate on bright
stars.  The special transient mode had the highest priority, and once this mode
was triggered by an alert, the regular SN or exoplanet survey would be
paused while the special observation was underway and resumed after it was done.

The {\it scheduler\/} keeps track of how many times a field has been
observed, and the last time it was observed, in the target-list configuration file.
Based on this information, and pre-defined rules of priority,
the scheduler updates the cadence for each field.  Moreover, if a field needs
a specific cadence, we can modify the configuration file and 
{\it scheduler\/} can easily take care of it.

In addition to the above scientific observing modes, there is a
separate mode where {\it scheduler\/} automatically switches to taking twilight flat-field
images when the Sun is 6--8.5\degr\ below the horizon.  The program is smart
enough to decide when to start and stop, what exposure times to use
based on the Sun's location, and which sky area has the
most uniform illumination based on local time and atmospheric
scattering models.

Based on measurements of the low sky background resulting from the clean air
and low dust and aerosol scattering at Dome~A,
the astronomical twilight has been redefined as  when 
the Sun is 13\degr\ below the horizon instead of the usual
18\degr\ \citep{Zou10}.  This gives us more than a month of
continuous dark time for observations in winter as well as partial
dark time in spring and autumn.

\subsection{Real-time Pipeline}
\label{sec:pipeline}

The AST3 data pipeline has also been designed for real-time transient discovery.
After various corrections to the images, aperture photometry
is applied to detect all sources in the images and to determine astrometric solutions.
Image-subtraction photometry (ISP) is then used to detect variable
sources accurately and efficiently (Sec.~\ref{sec:isp}).  All the
results are stored in a database on both {\it PIPE\/} and {\it MAIN\/} for
redundancy.

The real-time pipeline was originally designed for the SN search, but
it works satisfactorily for the special transient mode as well.  Given the science
requirements and the large amount of data
generated by the frequent exposures of the exoplanet search mode, no real-time
pipeline was originally specified, and it was planned to wait until the data were physically retrieved from the site.
However, as the project developed, some
data were required to be processed on-site and a separate pipeline was
later developed for the exoplanet search \citep{Zhang19a}.

\subsubsection{Image-Subtraction Photometry}
\label{sec:isp}

Following the basic raw CCD image processing (Sec.~\ref{sec:calib}), aperture
photometry was performed on all the scientific images with {\sc
sextractor} \citep{Bertin96}, and then astrometry with
{\sc scamp} \citep{Bertin06}, reaching a typical precision of 0\farcs1
\citep{Ma18}.

All detected sources were flux-calibrated against available catalogs in
the southern sky, such as the AAVSO Photometric All-Sky Survey
\citep{Henden16}.  Catalogs and light curves of the sources from
aperture photometry were generated, and these can be revised offline with
a more careful treatment once the data were retrieved from Dome~A
(\citealt{Ma18}; Yang et al. in preparation).  Here we focus on the real-time transient search with
image-subtraction photometry.  

The basic idea behind ISP is to subtract a reference image (the template)
from a science image of the same field so that variable sources can
be easily identified in the resulting difference image.  This is also called
difference image analysis.  A popular and efficient
software package for doing this is High-Order Transform of PSF and
Template Subtraction \citep[{\sc hotpants},][]{Becker15}, which
implements the algorithm developed by Alard and
Lupton \citep{Alard98,Alard00}.

Although {\sc hotpants} can cope with complicated issues of
different PSFs and spatially varying PSFs between the reference and
science images, we first need to prepare high quality reference images
for each survey field.  As the sky becomes gradually dark after
summer, the survey can cover more and more sky fields
until we can observe continuously for 24 hours every day during the polar
nights.  The first image of each field was taken as the initial reference
image, however, as the survey progressed we updated the references
as higher quality images -- in terms of sky
background, extinction, and PSF sharpness -- were obtained.
These images were registered and co-added to build the reference images
using {\sc swarp} \citep{Bertin02}.  All of this was done automatically
with supporting software within the AST3 pipeline on {\it {\it PIPE\/}}.

\begin{figure}
	\includegraphics[width=\columnwidth]{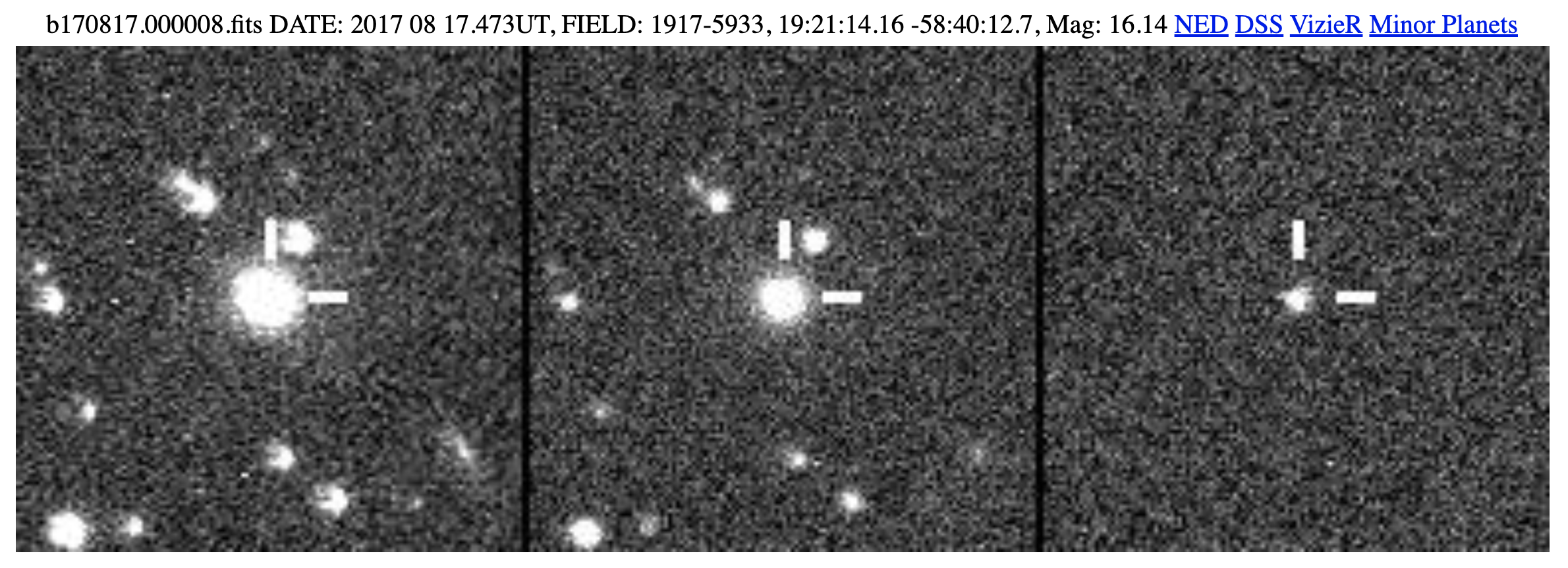}
    \caption{An example of transient stamp images on the internal web page of
real-time transient candidates \citep{Shang16}. From left to right: the reference
image, a new science image, and their difference.  Based on the
coordinates of each candidate, links to NED,
DSS, VizieR and Minor Planet Center are provided for easy
comparison.} 
    \label{fig:varcheck}
\end{figure}

Once {\sc hotpants} generates a difference image, {\sc sextractor} is
employed again to detect residual sources and do aperture photometry
to find transients.  Unfortunately, most of the residual
sources are false positives resulting from effects such as saturated stars, ghost images,
and detector defects.  We were able to discard most of the false positives by developing a series of criteria,
including the shape of sources (elongation, sharpness, and FWHM) and
the position of sources (whether the near edges of an image, defects of the
CCD, or saturated stars).
Finally we assign a score for each remaining transient candidate based
on its likelihood of being a real source, and whether it is close to
a known galaxy, which increases the chance of it being a SN.  Balancing these two factors, the
position, light curve, and stamp images of the most likely SN and
transients were transferred back to NAOC and put on an internal web page for further
visual inspection \citep{Shang16} and for decisions on follow-up observations with other
telescopes
(Fig.~\ref{fig:varcheck}). 

\subsubsection{Special Correction Requirements}
\label{sec:calib}

The usual raw CCD image processing involves corrections for overscan, dark current, and
flat-fields.  However, given some special difficulties with
AST3 operating at Dome~A, we had to develop new methods for
flat-fielding and dark corrections.  These had to be done carefully and
interactively after data were retrieved from Dome~A, therefore, high
quality flat-field and dark images were not available for the very
first year.

Twilight flats were taken at the beginning of each observing season.
Since the twilight sky is not
very flat in AST3's large FOV, we have to remove the slope in the images
before combining them into a master flat-field frame \citep{Wei14}.

The dark current correction was a little more complicated. Since AST3
does not have a shutter, we cannot take dark frames on-site and the
dark frames taken previously in the laboratory could not be used since the pattern
had changed.  We were forced to develop a new method to derive a dark
frame from scientific images and apply it to dark correction
\citep{Ma14b,Ma18}.  This turned out to be very efficient and greatly
improved our photometric accuracy.

Electrical cross-talk was observed between different readout channels when one channel 
is reading out a saturated star.  This needs to be removed early in the processing.
However, since this effect was complicated,
it was studied and removed as part of the
offline aperture photometry pipeline which is more complete and
results in better photometric accuracy \citep{Ma18}.  
This correction was added to the real-time on-site pipeline after the
first season.

\begin{figure}
	\includegraphics[width=\columnwidth]{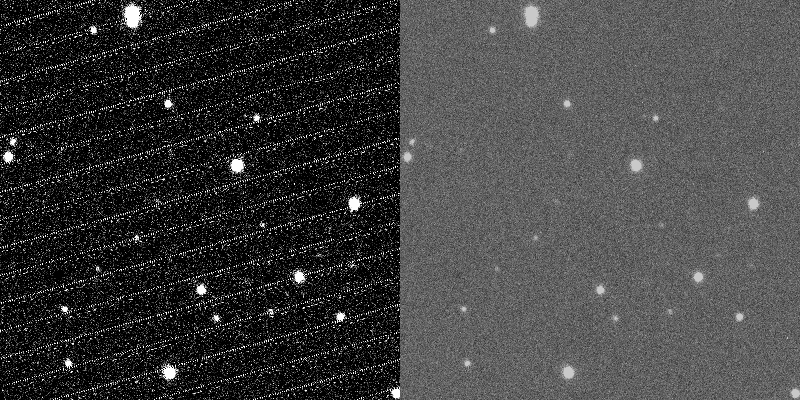}
    \caption{Left panel: an example of electromagnetic interference in the raw images from AST3-2 during 2017 due to a break in a cable shield. Right panel: the same image after removing the noise, using the method discussed in the text.} 
    \label{fig:noise}
\end{figure}

A final problem that affected the raw images from 2017 was the presence in the raw images of
diagonal stripes, see Fig.~\ref{fig:noise}. Investigation showed that the noise frequency was 16 kHz, the same as the chopping frequency for the telescope DC motor drives. Subsequent inspection during the next summer at Dome~A showed a broken cable shield. Fortunately, the noise could be almost entirely removed, as shown in the right panel of Fig.~\ref{fig:noise}, by subtracting from each of the 16 CCD channels an appropriately filtered image from the signals from the other 15 channels. The noise disappeared following repair of the cable shield in January 2019.

\subsection{Interaction Between NAOC and Dome~A}
\label{sec:alarm}

While the telescope and associated systems were designed for completely automated operation, there were times during the commissioning phase when we had to run interactively. There were also some periods when instruments did not work as expected, and we had to operate the telescope
remotely and manually 24
hours a day.  The situation
has improved over the years.

The {\it MAIN\/} computer
monitors all the instruments and devices and collects their status
information, which is sent back to NAOC every 10 minutes and displayed
on an internal web page for personnel to check \citep{Shang16}.  
If there are cases where unexpected problems with instruments or devices are
detected, besides shutting things down as necessary, the alarm system
on {\it MAIN\/} will send an emergency message back to NAOC, and the
message is
immediately forwarded as an SMS to people on duty to remotely login to
{\it MAIN\/}, check and take care of the problem.

Since the bandwidth over Iridium is limited, expensive, and
sometimes unstable, in order to improve the data transfer efficiency
and reliability, we have developed and implemented our own narrow-band
file transfer protocol (NBFTP).   This takes into account our specific
transmission requirements and is able to save up to 60\% of data
consumption for small files (Huang et al. in preparation).

\section{Summary of the AST3 Automated Sky Survey }
\label{sec:summary}

\begin{figure}
	\includegraphics[width=\columnwidth]{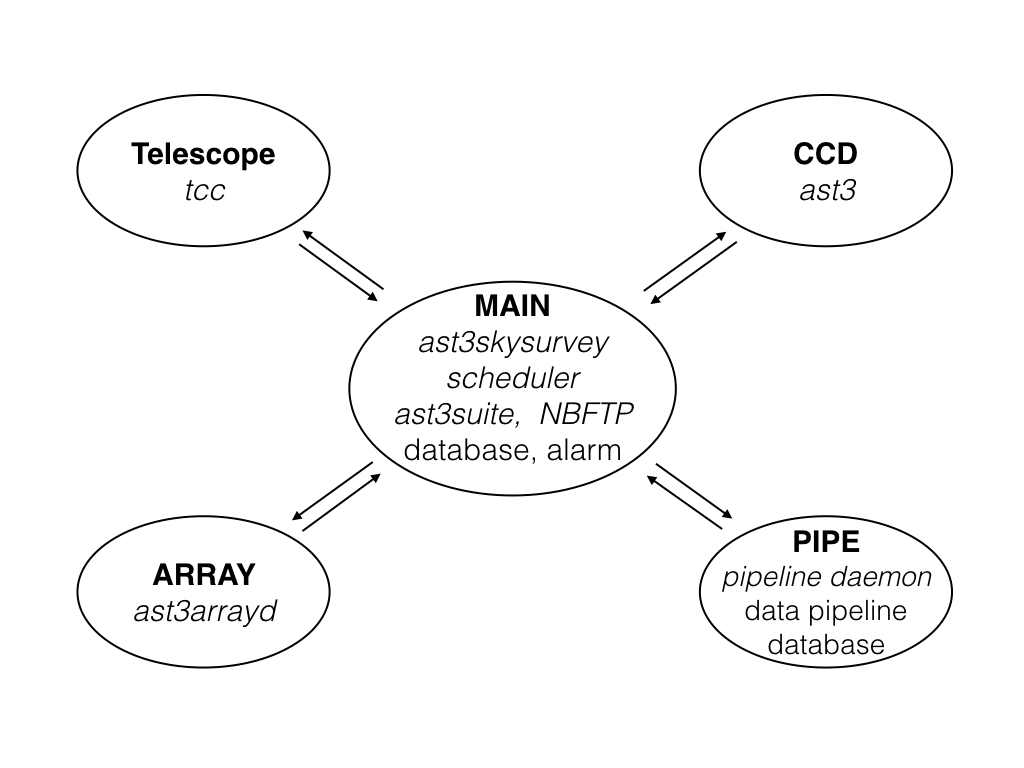}
    \caption{Logic diagram of the key components of the sky survey and their interactions. Those names in italic are the software programs.}
    \label{fig:main}
\end{figure}

The unattended AST3 sky survey has been fully automated from
scheduling of the observations to posting of transient candidate alerts.
Figure~\ref{fig:main} outlines key components of the sky survey, and their interactions.

{\it MAIN\/} is the core computer, and the survey program {\it ast3skysurvey\/}
running on {\it MAIN\/} performs all the tasks.  It is a script-based
program managing and monitoring the entire system, calling basic
commands (tasks) of {\it ast3suite\/} and communicating with various
network ports, to fulfill the automated survey requirements. 
During the continuous survey, many things are done in parallel,
including image readout from the CCD, re-pointing telescope, data storage,
and data reduction (pipeline), greatly increasing the survey
efficiency.

The survey usually starts in March each year, when we start {\it ast3skysurvey\/}, which asks {\it scheduler\/}
for the next observation.  When there is an available target to observe, either for
the scientific survey or a twilight flat, {\it scheduler\/} will return coordinates
for
{\it ast3skysurvey\/} to call the telescope to move, and then the
CCD camera to take exposures.  

The program {\it tcc\/} running on the {\it TCC\/} computer drives
the telescope.  {\it Ast3skysurvey\/} communicates with {\it tcc\/} based
on a customized command set built on TCP/IP. The set has about 50
commands that control the telescope’s pointing, tracking, and
configuration. The commands also monitor and record the telescope’s status, and
adjust various parameters such as those of the telescope motor drives.

Once an exposure has been completed, the image is rapidly frame-transferred to the
buffer regions of the CCD for subsequent slower readout.  During the readout period {\it
ast3skysurvey\/} can ask {\it scheduler\/} for a new target and start to move
the telescope, and the next loop
begins (Fig.~\ref{fig:ast3suite}).  So in this frame-transfer mode, 
we could, in principle,
move the telescope during the readout time.  However, in practice
this seldom happens since we usually take several
exposures for each field, so the telescope cannot be moved until they are all complete.

After an image is read out, while the survey itself continues, {\it
MAIN\/} (through {\it ast3suite\/}) adds various items of information such as the 
time, coordinates on the sky, telescope and CCD status, and weather conditions, to the image FITS headers for future reference.  It then sends the
raw image simultaneously to {\it ARRAY\/} for permanent storage and to
{\it PIPE\/} for data reduction.

As soon as {\it PIPE\/} receives an image, the {\it Pipeline Daemon\/}
triggers the pipeline program which processes the image through the basic CCD
corrections, followed by aperture photometry, astrometry, and ISP. As discussed in
Sec.~\ref{sec:isp}, the ISP reference templates are iteratively improved.  Following ISP, a second run of
aperture photometry is performed on the difference image, detecting all
possible sources.  Based on pre-defined criteria, false positives are
rejected and transient candidates, if any, merged.  The top
candidates are sent back to NAOC every 0.5\,hour to be displayed on a
web page for further confirmation and decisions on follow-up
spectroscopy observations (Fig.~\ref{fig:varcheck}).

AST3 ISP identifies variable sources efficiently and has discovered
SNe \citep{Ma14c,Ma14d,Uddin17}, dwarf novae \citep{Ma16}, minor
planets, AGNs, and numerous variable stars, although we
do not report variable stars in real-time since they are found and measured with an
offline pipeline later \citep{Ma18,Wang17}.  Due to some technical
issues, the limiting magnitude of the AST3 survey could not go as deep as originally expected
for an efficient SN survey (\citealt{Ma18}; Yang et al. in preparation), however, the exoplanet
search has turned out to be very successful with 116 transiting
exoplanet candidates being found in one season \citep{Zhang19b}.

Although the survey is designed for full automation, each year before
the observing season we manually operate the telescope and instruments to
make sure that everything functions properly.  Also, in case of a malfunction,
{\it ast3skysurvey\/} sends alarms from Dome~A and we have to interrupt
the survey manually, whether or not the problem is serious enough
for the system to stop by itself to protect the hardware (Sec.~\ref{sec:alarm}).

In 2017 AST3-2
was able to operate through the whole winter and also contributed to
observations and measurements of the first optical counterpart of the
gravitational wave source GW170817 \citep{Abbott17,Hulei17,Andreoni17}.

\section*{Acknowledgments}

The authors deeply appreciate the CHINARE for their continuous support
of astronomy at Dome~A.  This study has been supported by the
National Basic Research Program (973 Program) of China (Grant No.\
2013CB834900), the Chinese Polar Environment Comprehensive
Investigation $\&$ Assessment Program (Grant No.\ CHINARE2016-02-03),
and the National Natural Science Foundation of China (NSFC) (Grant
Nos.\ 11003027, 11203039, 11273019, 11403057, 11403048 and 11733007). PLATO-A
is supported by a grant from the Australian Antarctic Division, and the NCRIS
scheme administered by Astronomy Australia Limited.








%
%


\bsp	
\label{lastpage}
\end{document}